\documentclass[a4paper,english]{article}
\usepackage{lmodern}

\usepackage[T1]{fontenc}
\usepackage[latin9]{inputenc}
\usepackage{graphicx}
\usepackage{amssymb}
\usepackage{nomencl}
\providecommand{\printnomenclature}{\printglossary}
\providecommand{\makenomenclature}{\makeglossary}
\makenomenclature

\makeatletter
\date{}

\usepackage{babel}

\makeatother

\makeatother

\usepackage{babel}

\begin{document}

\title{Ion Beam Shepherd for Asteroid Deflection}


\author{{\normalsize Claudio Bombardelli}%
\thanks{Research Fellow, ETSI Aeronauticos, Plaza Cardenal Cisneros 3%
} {\normalsize and Jesus Peláez}%
\thanks{Professor, ETSI Aeronauticos, Plaza Cardenal Cisneros 3, AIAA Member%
} \\
 \textit{\normalsize Technical University of Madrid (UPM), Madrid,
E-28040,}{\normalsize {} Spain}\\
 \\
 {\normalsize {} }}

\maketitle
\begin{center}
(Accepted for publication. Journal of Guidance, Control and Dynamics.
January 2011. Submitted July 2010)
\par\end{center}

\printnomenclature{}


\section{Introduction}

$\qquad$Asteroid deflection is becoming a key topic in astrodynamics.
Although no asteroid has been deflected so far, altering the trajectory
of a small-size asteroid to avoid a catastrophic impact with the Earth
has been shown to be, in principle, technically feasible \cite{ahrens1992deflection}
and different techniques, ranging from nuclear detonation to kinetic
impact and low-thrust methods, have been proposed \cite{ahrens1992deflection},\cite{melosh1994non},\cite{scheeres2004mechanics}.
Each of these methods shows advantages and drawbacks that in general
depend on the mass and orbital characteristics of the particular asteroid
to be deflected as well as its physical property (porosity, composition,
surface reflectivity, etc.) and rotation state.

Among the low-thrust methods, in which the asteroid trajectory is
altered by a small and continuous push, a very interesting solution
was proposed in 2005 by Lu and Love\cite{lu2005gravitational}. The
method, named {}``gravity tractor'' or {}``gravity tugboat'' exploits
the gravitational interaction between an Earth-threatening asteroid
and a spacecraft hovering above its surface to achieve a contactless
deflection of the former. In the above article, it was shown that
a 20-tonne spacecraft could deflect a typical asteroid of about 200-m
diameter within one year of hovering time and given a lead time of
20 years. The possibility to predictably change the asteroid orbit
with no need of physical attachment and irrespectively of the mechanical
properties of the asteroid makes the gravity tractor concept one of
the preferred deflection strategies for sub-kilometer asteroids whose
orbit characteristics are known with sufficient time before the predicted
impact. In addition, the gravity tractor can be used in conjunction
to less accurate deflection methods (e.g. kinetic impactor or nuclear
detonation) to provide a deflection trim capability and avoid secondary
impacts. 

However, while offering the undoubtable advantage of no physical attachment
with the asteroid the gravity tractor concept suffers from at least
two major drawbacks.

The first is the need for a massive spacecraft to physically produce
the gravitational force required to slowly deflect the asteroid. As
it will be shown in this article, in order to achieve a given gravitational
pull a gravity tractor needs to carry a total mass which greatly exceeds
the mass required (in terms of propellant and power system mass) to
counteract such force with an optimized electric propulsion system.
While the extra mass can be used for other spacecraft functions (e.g.
scientific payloads), the need to deliver it up to the asteroid orbit
will affect the total mission cost significantly.

The second is the need for a continuous control of the spacecraft
hovering distance, which has to be fairly small (a fraction of the
asteroid radius) if sufficient force is to be achieved. The instability
associated with the hovering equilibrium position and the rotation
of the generally irregularly-shaped asteroid poses collision risks
and complicates the matter even further.

Recently, these authors have proposed a new propulsion concept \cite{bombardelli_patent}
in which a highly collimated, high-velocity ion beam is produced by
an ion thruster on board a shepherd spacecraft and pointed against
a target to modify its orbit and/or attitude with no need for docking.
If the ion beam is correctly pointed at the target the momentum transmitted
(ions have been accelerated up to 30 km/s and more on board spacecraft
in past missions) can reach the same magnitude that would be obtained
if the target object had the same ion thruster mounted on its own
structure. The same concept can be advantageously applied to the contactless
deorbiting of space debris in low Earth orbit \cite{bombardelli2010debris}
and Earth geostationary orbit \cite{Kitamura_debris}, a theme that
is gaining considerable interest in the field of space science and
utilization. Note that the idea of accelerating a spacecraft with
a flux of incident ions was also recently explored by Brown et al.
\cite{brown2007lunar} who propose a lunar-based ion-beam generator
to remotely propel spacecraft in the Earth-Moon system. 

As it will be shown in this article, this concept can be used to alter
the trajectory of earth threatening asteroids with a much higher efficiency
when compared to the gravity tractor concept.  \\


\section{Ion Beam Shepherd Satellite}

\begin{figure}[!t]
\centerline{\includegraphics[width=10cm]{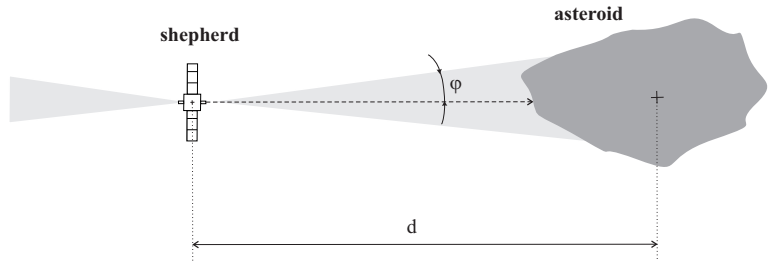}}

\caption{\label{fig:fig1}Schematic of asteroid deflection with an Ion beam
shepherd. The shepherd spacecraft directs a stream of accelerated
ions towards the asteroid surface so that the momentum carried by
the stream is transferred to the asteroid to be deflected. A second
ion prupulsion unit thrusting in the opposite direction is needed
in order to keep a constant distance between the shepherd and the
asteroid.}

\end{figure}

The concept of Ion Beam Shepherd (IBS) applied to asteroid deflection
is schematized in Fig.\ref{fig:fig1}. The shepherd spacecraft is
located not too far from the asteroid and pointing one of its ion
thrusters directly at the asteroid surface. The high-velocity ions
of the quasi-neutral%
\footnote{As it is always the case in electric propulsion technology the plasma
leaving the propulsion system is neutralized in order to avoid a net
charge to accumulate on the spacecraft %
} plasma emitted by the thruster reach the asteroid surface transmitting
their momentum. Assuming the collision is predominantly inelastic
and that the beam fully intercepts the surface of the asteroid, the
latter will undergo a force roughly equal and opposite to the one
experienced by the spacecraft. It will then be necessary to have a
second ion thruster mounted on the spacecraft to cancel out the total
force and keep constant the distance with respect to the asteroid.

In the real case, secondary ions and neutrals are sputtered back from
the surface increasing, in principle, the net momentum transmitted
to the asteroid. Yet their ejection velocities are generally small
compared to the ones of the impinging ions \cite{SRIM} so that in
the end the effect on the transmitted force is negligible. On the
other hand, a decrease in the total transmitted momentum occurs when
part of the ions miss the target due to ion beam divergence effects
and possible beam pointing errors. In order for the beam to fully
intercept the asteroid surface the hovering distance of the spacecraft
must not exceed the value:

\[
d_{max}\backsimeq\frac{s}{2\sin\varphi},\]

where $s$ denotes the smaller asteroid dimension and $\varphi$ is
the divergence angle of the beam (Figure \ref{fig:fig1}).

State of the art ion thruster can reach half-cone divergence angles
as low as 15 degrees \cite{gardner97predictions} which would allow,
for example, to fully intercept a spherical asteroid at a distance
of about twice its diameter. At such distances the risk of collision
is greatly reduced when compared to the case of a closely hovering
gravity tractor. At the same time, as it will be shown later in the
article, the resulting gravitational pull would be negligible, compared
with the force provided by the thruster, hence greatly reducing the
instability of the hovering motion, but still large enough to be detected
by onboard measurement systems so that it can be used for estimating
the distance between the asteroid and the spacecraft.

If one assumes, for simplicity, that the asteroid orbit is quasi-circular,
the force needed to produce a velocity change $\Delta V$ after a
hovering time $\Delta t$ for a spherical asteroid of diameter $d_{ast}$
and density $\rho$ is \cite{lu2005gravitational}:

\[
F_{th}=2\times\frac{\Delta V}{\Delta t}\times\frac{4}{3}\rho\pi(d_{ast}/2)^{3},\]

The total propellant mass spent after the hovering time $\Delta t$
is:

\[
m_{fuel}=\frac{2F_{th}\Delta t}{v_{E}}\]

where $v_{E}$ is the ion ejection velocity and the initial factor
of 2 takes into account the need for a second thruster to bring to
zero the net thrust force on the spacecraft%
\footnote{The additional force needed to deflect the beam shepherd satellite
from its orginal orbit (by the same amount as the asteroid) is clearly
negligible %
}. 

The mass of the spacecraft power plant needed to produce the force
$2F_{th}$ is:

\[
m_{pp}=\frac{2F_{th}\alpha v_{E}}{2\eta},\]

where $\eta$ is the thruster efficiency and $\alpha$ the inverse
specific power (kg/W) of the power plant feeding the electric propulsion
system. 

The total spacecraft mass needed to accomplish the deflection is obtained
by adding to the latter two terms the structure mass $m_{str}$:

\[
m_{IBS}=m_{fuel}+m_{pp}+m_{str}=\frac{\pi\rho d_{ast}^{3}}{6}\Delta V\left(\frac{2}{v_{E}}+\frac{\alpha v_{E}}{\eta\Delta t}\right)+m_{str}\]

The latter equation can be used to find the optimum value of the ion
thruster exhaust velocity which turns out to be the Irving-Stuhlinger%
\footnote{Note that in Stuhlinger book the thruster efficiency is not accounted
for in the formula and that the specific power, rather than the inverse
specific power, is employed%
} characteristic velocity\cite{stuhlinger1964ion}:

\[
v_{E}^{opt}=\sqrt{\frac{2\eta\Delta t}{\alpha}}\]

The IBS mass has to be compared with the one of a gravity tractor
achieving the same deflection $\Delta V$ after a hovering time $\Delta t$
which is \cite{lu2005gravitational}:

\[
m_{GT}=\frac{\Delta V\:\left(kd_{ast}/2\right)^{2}}{G\Delta t}\]

where $G$ is the gravitational constant and $k$ is the hovering
distance measured in asteroid radii.

\begin{figure}[!t]
\centerline{\includegraphics[width=8cm]{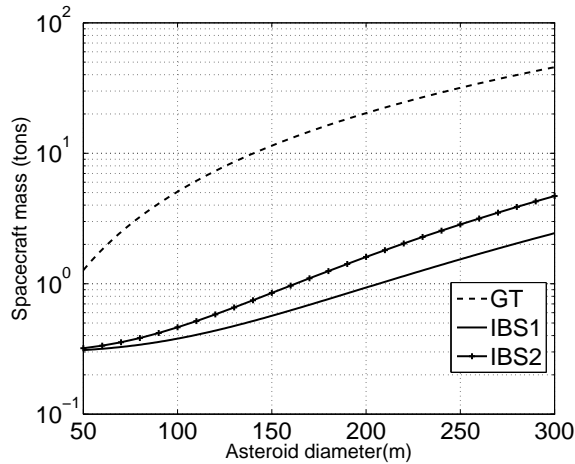}}

\caption{\label{fig:fig2}Total spacecraft mass required for deflecting asteroids
of different size using the gravity tractor vs ion beam shepherd approaches.
The deflection velocity change is set to $1.9\times10^{-3}ms^{-1}$
after a hovering time of one year. The asteroid is assumed spherical
with mass density of 2000 $\mathrm{kg\, m^{-3}}$\cite{lu2005gravitational}.
The gravity tractor is kept at constant hovering distance equal to
1.5 asteroid radii. Two different configurations of IBS spacecraft
are considered. A {}``near-future'' design (IBS1) employs an ion
thruster with 80\% thrust efficiency and 10000 s specific impulse
($v_{E}\sim100\mathrm{\: km/s}$) and a power plant with $\alpha=5\mathrm{\: kg/kW}$.
A {}``state-of-the-art'' design (IBS2) employs an ion thruster with
60\% thrust efficiency and 3100 s specific impulse ($v_{E}\sim30\mathrm{\: km/s}$)
and $\alpha=10\mathrm{\: kg/kW}$. Structural mass is set, as a preliminary
value, to 300 kg in both cases.}

\end{figure}

A comparison of the total mass required to deflect asteroids of different
diameters using the IBS and the gravity tractor concept is shown in
Figure \ref{fig:fig2} in which the two deflection systems provide
during one year a velocity change of $1.9\times10^{-3}ms^{-1}$, which
is enough to deflect a typical asteroid of 200 m given a 20 years
lead time\cite{lu2005gravitational}. The IBS allows more than one
order of magnitude mass savings when compared to the gravity tractor
concept with the difference increasing the smaller the asteroid diameter. 

Two different propulsion systems are considered for the IBS concept:
an advanced propulsion and power system (IBS1) with 80\% efficiency
and providing 10000 s specific impulse (which corresponds to the optimum
value for a thrust time of one year) and 5 kW/kg power density, as
well as a state-of-the-art system\cite{brophy2000ion} (IBS2) with
60\% efficiency, 3100-s specific impulse and 10 kW/kg power density.
Note that assuming a constant power density is correct if a nuclear
power plant is employed or if, in case solar power is employed, the
asteroid orbit is nearly circular.

In particular, the deflection of a 200-m diameter asteroid, which
would require a 20-tonne gravity tractor, can be accomplished with
an ion-beam shepherd spacecraft weighting less than 1 tonne employing
high-efficiency and high-specific impulse ion thrusters available
in the near future, or less than 2 tonne with state-of-the-art hardware.

\begin{figure}[!t]
\centerline{\includegraphics[width=8cm]{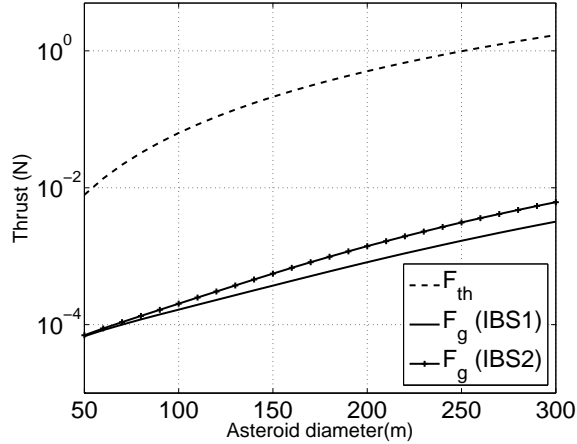}}

\caption{\label{fig:fig3}Comparison between the thrust force $F_{th}$ exherted
on the IBS by the asteroid and the mutual gravitational force $F_{g}$
as a function of the asteroid diameter. The deflection magnitude,
asteroid density and IBS design are the same as in Fig. \ref{fig:fig2}.
A hovering distance of two asteroid diameters from the center is assumed.}

\end{figure}

\begin{figure}[!t]
\centerline{\includegraphics[width=7cm]{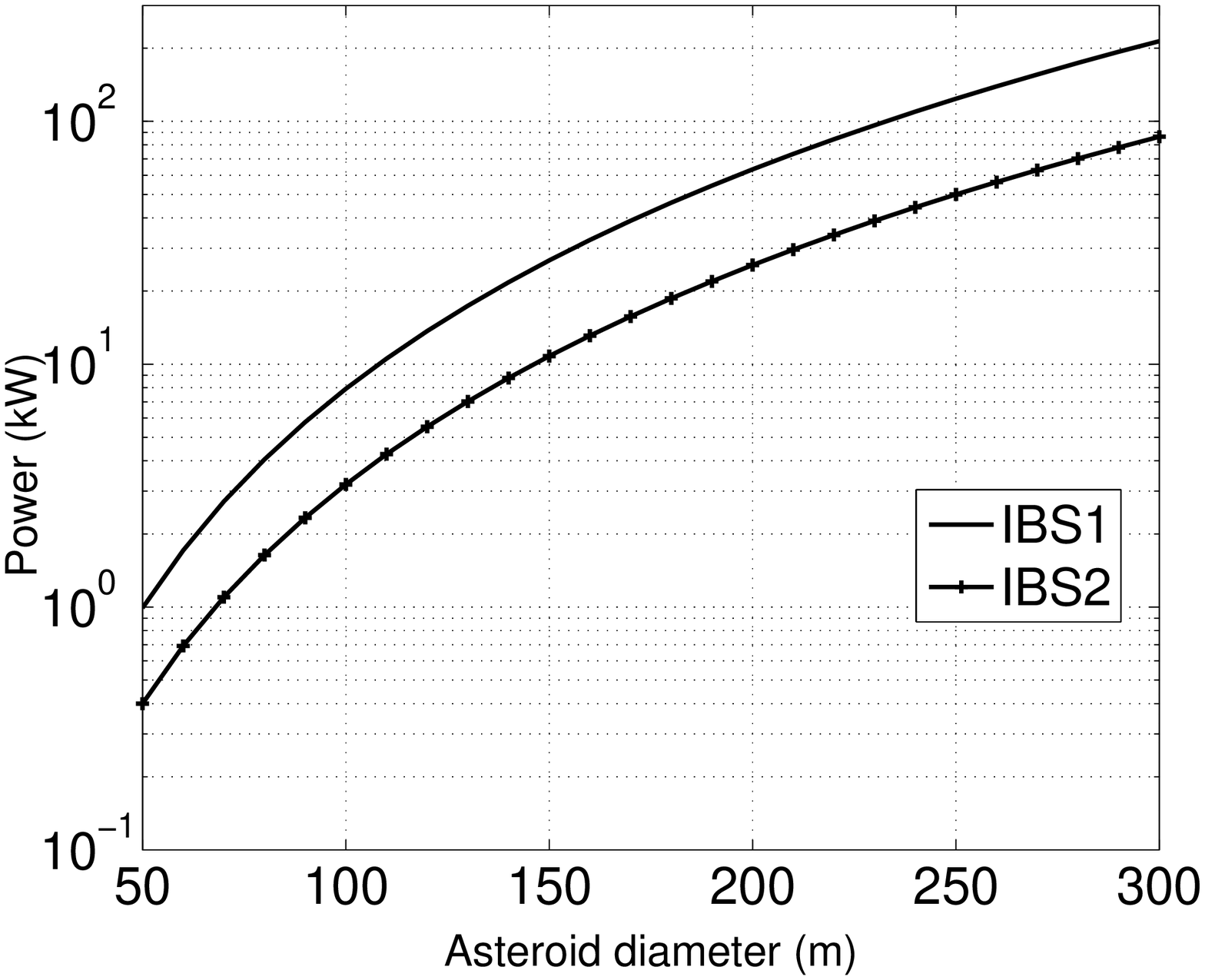}\includegraphics[width=7cm]{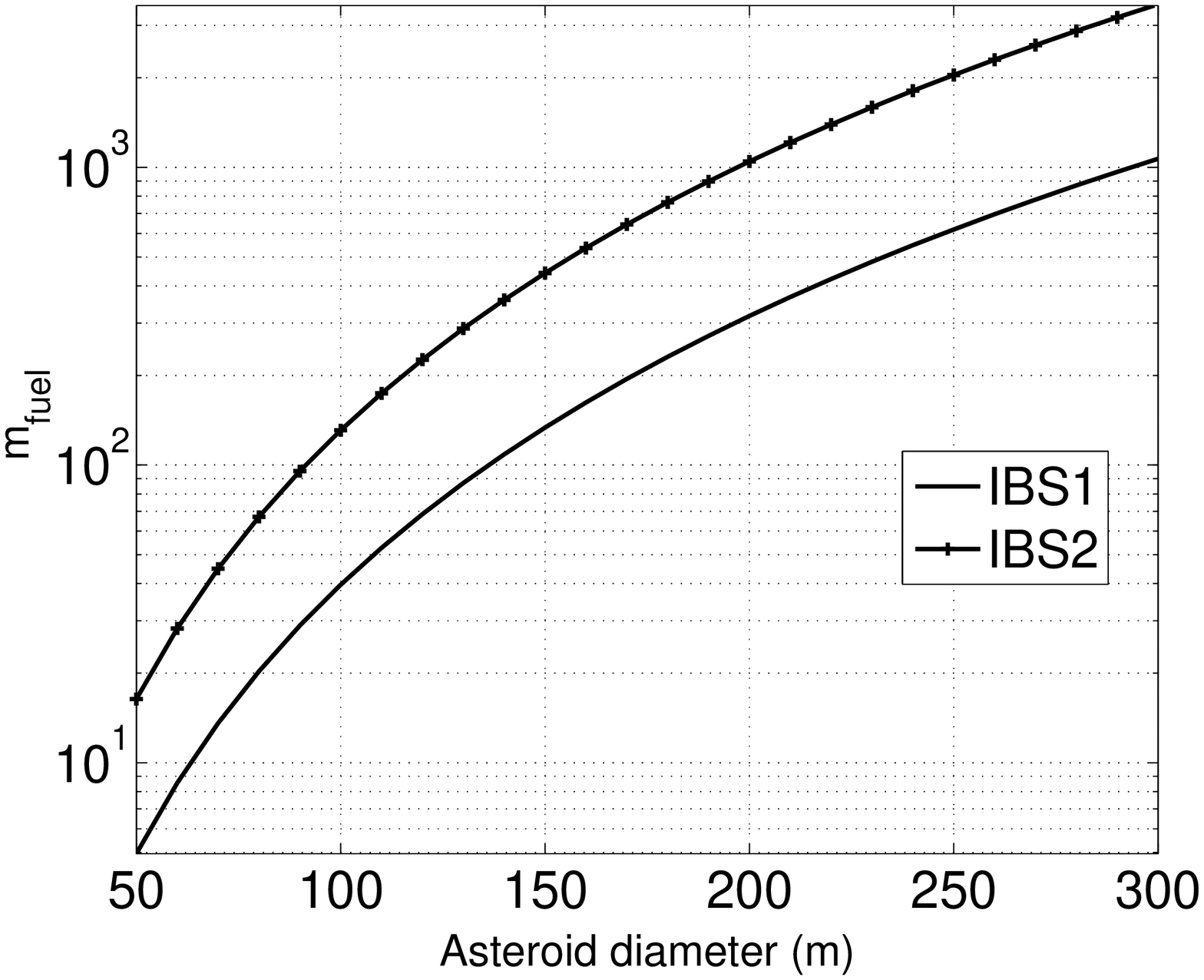}}

\caption{\label{fig:fig4}Power and total propellant consumption for the two
IBS designs considered in Fig. \ref{fig:fig2}and using the same asteroid
deflection magnitude and density. }

\end{figure}

Additional plots (Fig. \ref{fig:fig3},\ref{fig:fig4} ) compare the
value of the ion beam thrust force on the asteroid with the (negligible)
mutual gravitational attraction between the latter and an IBS spacecraft
hovering at two asteroid diameters from the center (providing full
beam interception with a 15 degree half-cone divergence) and provide
the values for the power level and total propellant consumption throughout
the mission.

\section{Conclusions and Recommendations}

A new concept for low-thrust asteroid deflection has been presented
that exploits the momentum transmitted by a low-divergence accelerated
ion beam flux from the propulsion system of a nearby spacecraft. The
concept, which shares with the gravity tractor the ability of deflecting
an asteroid without any physical contact, allows more than one order
of magnitude mass savings when compared with the former and does not
require close hovering hence greatly simplifying the spacecraft control
problem. Given these improvements and because low-divergence ion beams
are routinely employed in spacecraft technology, an asteroid deflection
demonstration mission may be in reach in the near future. Future studies
will be needed to evaluate the actual deflection performance of the
system for different asteroid orbits and to compare it with other
short-term deflection methods such as the kinetic impactor. \\

\section{Acknowledgments}

The work for this paper was supported by the research project \textquotedblleft{}Simulación
Dinámica de Sistemas Espaciales Complejos\textquotedblright{} supported
by the Dirección General de Investigación (DGI) of the Spanish Ministry
of Education and Science through the contract AYA2010-18796 and by
the {}``ARIADNA Call for Ideas on Active Debris Removal'', established
by the Advanced Concepts Team of the European Space Agency.

\medskip{}

\bibliographystyle{aiaa}
\bibliography{library}

\begin{thebibliography}{10}
\newcommand{\enquote}[1]{``#1''}

\bibitem{ahrens1992deflection}
Ahrens, T. and Harris, A., \enquote{{Deflection and fragmentation of near-Earth
  asteroids},} {\em Nature\/}, Vol.~360, No. 6403, 1992, pp.~429--433.

\bibitem{melosh1994non}
Melosh, H., Nemchinov, I., and Zetzer, Y., \enquote{{Non-nuclear strategies for
  deflecting comets and asteroids.}} {\em Hazards due to comets and
  asteroids\/}, Vol.~1, 1994, pp. 1111--1132.

\bibitem{scheeres2004mechanics}
Scheeres, D. and Schweickart, R., \enquote{{The mechanics of moving
  asteroids},} {\em 2004 Planetary Defense Conference: Protecting Earth from
  Asteroids\/}, 2004.

\bibitem{lu2005gravitational}
Lu, E. and Love, S., \enquote{{Gravitational tractor for towing asteroids},}
  {\em Nature\/}, Vol.~438, No. 7065, 2005, pp.~177--178.

\bibitem{bombardelli_patent}
Bombardelli, C. and Pelaez, J., \enquote{{Sistema de modificaci\'{o}n de la
  posici\'{o}n y actitud de cuerpos en \'{o}rbita por medio de sat\'{e}lites
  gu\'{\i}a},} Patent number P201030354. Presented at the Spanish Patent Office
  on March 11, 2010. PCT Patent Application PCT/ES2011/000011.

\bibitem{bombardelli2010debris}
Bombardelli, C. and Pelaez, J., \enquote{{Ion beam shepherd for contactless
  space debris removal},} Journal of Guidance, Control and Dynamics. In press.

\bibitem{Kitamura_debris}
Kitamura, S., \enquote{{Large Space Debris Reorbiter using Ion Beam
  Irradiation},} {\em 61 st International Astronautical Congress, Prague, Czech
  Republic\/}, Paper IAC-10.A6.4.8, September 2010.

\bibitem{brown2007lunar}
Brown, I., Lane, J., and Youngquist, R., \enquote{{A lunar-based spacecraft
  propulsion concept--The ion beam sail},} {\em Acta Astronautica\/}, Vol.~60,
  No. 10-11, 2007, pp.~834--845.

\bibitem{SRIM}
J.~F.~Ziegler, J. P.~B. and Ziegler, M.~D., {\em SRIM - The Stopping and Range
  of Ions in Matter\/}, Lulu Press, 2007.

\bibitem{gardner97predictions}
Gardner, B., Katz, I., and Briuza, D., \enquote{{Predictions of NSTAR charge
  exchange ions and contamination backflow},} {\em Proceedings of the
  International Electric Propulsion Conference. Cleveland OH, August 24-28
  1997\/}, Vol.~1, pp. 281--289.

\bibitem{stuhlinger1964ion}
Stuhlinger, E., {\em {Ion propulsion for space flight}\/}, McGraw-Hill New
  York, 1964, chapter 4.1.

\bibitem{brophy2000ion}
Brophy, J., Kakuda, R., Polk, J., Anderson, J., Sovey, J., Patterson, M., Bond,
  T., Christensen, J., Matranga, M., and Bushway, D., \enquote{{Ion Propulsion
  System(NSTAR) DS 1 technology validation report},} {\em Deep Space 1
  technology validation reports(A 01-26126 06-12), Pasadena, CA, Jet Propulsion
  Laboratory(JPL Publication 00-10)\/}, 2000.

\end{thebibliography}

\end{document}